**Exponential Pixelating Integral Transform with Dual Fractal Features for Enhanced Chest X-Ray Abnormality Detection**

Naveenraj K[a], Sri Ram Macharla[b], Kanimozhi M[c], Sudhakar M S [c*]

a – The Construct, Barcelona, Spain

b – Involgixs Inc, USA

c - School of Electronics Engineering, Vellore Institute of Technology, Vellore, Tamilnadu, India

* - Corresponding Author

**Abstract**

The heightened prevalence of Respiratory Disorders (RD), particularly exacerbated by a significant upswing in fatalities due to the novel coronavirus, underscores the critical need for early detection and timely intervention. This imperative is paramount, possessing the potential to profoundly impact and safeguard numerous lives. Medically, chest radiography stands out as an essential and economically viable medical imaging approach for diagnosing and assessing the severity of diverse RD. However, their detection in Chest X-rays (CXR) is a cumbersome task even for well-trained radiologists owing to low contrast issues, overlapping of the tissue structures, subjective variability, and the presence of noise. To address these issues, a novel analytical model termed Exponential Pixelating Integral (EPI) is introduced for the automatic detection of infections in CXRs in this work. Initially, the presented EPI enhances the pixel intensities to overcome the low-contrast issues that are then polar-transformed followed by their representation using the locally invariant Mandelbrot and Julia fractal geometries for effective distinction of structural features. The collated features labeled Exponential Pixelating Integral with dually characterized Fractal features (EPIFF) are then classified by the non-parametric Multivariate Adaptive Regression Splines (MARS) to establish an ensemble model between each pair of classes for effective diagnosis of diverse Respiratory Disorders. Rigorous analysis of the proposed diagnostic framework on large medical benchmarked datasets showcases its superiority over its peers by registering a higher classification accuracy and F1 scores ranging from 98.46-99.45% and 96.53-98.10% respectively, making it a precise and interpretable automated system for diagnosing diseases.

## 1. Introduction

Globally Respiratory Disorders are the major cause of death, according to the World Health Organization. Especially, Edema, Effusion, Emphysema, COVID-19, and Pneumonia are the most common lung diseases found in both children and adults due to the infection caused by viruses, bacteria, and fungi [1]. Pulmonary edema occurs because of the accumulation of excess fluid in the alveoli of the lungs leading to shortness of breath and ending in death if untreated. Likewise, Effusion is caused by the buildup of excess fluid between the layers of the pleura outside the lungs and poses a setback in distinguishing from edema in CXRs. While the progressive lung disease emphysema leads to the gradual deterioration of lung tissue, particularly the alveoli, it culminates in the bursting of air sacs with a simultaneous reduction of lung surface area, obstructing breathing. In addition to these conditions, the infection of the lungs by bacteria, fungi, and viruses, such as pneumonia, results in inflammation and fluid accumulation. Likewise, COVID-19 induced by Syndrome Coronavirus 2 (SARS-CoV-2) virally infects the lungs or pneumonia affecting the upper respiratory tract. Apart from viral pneumonia, bacterial pneumonia caused by pathogens such as Streptococcus leads to lung inflammation[2].

According to recent studies, around 500 million are diagnosed with diverse pneumonia, and nearly 1.5 million die each year [4]. Also, the contagious nature of a few of these ailments requires prompt diagnosis and isolation, with varying treatment options[3]. Therefore, early identification and intervention will greatly improve the course of the illness, slowing progression, reducing symptoms, and decreasing the frequency of exacerbations [3,5,6]. Over the years, CXRs' and Computed Tomography (CT) modalities have played an important role in acutely diagnosing such ailments by assessing their severity through fluid presence, lung space, and other lesions. While the CXR is a fast imaging, less sensitive, and cost-effective modality than the CT, thereby, making it widely popular in the identification of common respiratory diseases, especially in low-resource hospitals[7,8].

However, CXRs are difficult to interpret because of anatomy ending in structural superimposition [9]. The natural pattern of branching blood arteries in the lung fields makes it difficult for even the most skilled radiologists to tell infiltrates apart from normal tissue, and minuscule nodules hinder lesion identification [10] and manual detection is time-consuming which are substantial variations in severity assessments. In addition, a lack of expertise contributes to a high rate of inaccurate assessments. These issues highlight the clinical significance and complexity of chest radiographs thereby, motivating research on automated diagnostic techniques to aid radiologists in image interpretation. Also, diagnostic efficiency, precision, and speed can be significantly enhanced by automating these processes on a large scale, and accessible to all.

As a result, more precise and automated interpretation methods for diagnosing RDs are desired and are broadly categorized as conventional handcrafted or Deep Learning (DL) models, depending on the choice of feature acquisition. Numerous investigations have demonstrated that convolution neural networks (CNN) are superior and have led to an increase in their utilization for chest diagnosis [3,11–16]. In particular, the Deep CNN [17]**,** utilized fractional-order derivative to extract the textural details of the ROC region from CXRs, which are then fed to a gray relational analysis for diagnosis of diverse RDs. Similarly, [18] determined the best convolution architecture using the learning-by-training mechanism that effectively detected pneumonia in CXRs. However, the accuracy and reliability of models depended heavily on the labels given by humans and on the choice of hyperparameters. On the other hand, Transfer learning successfully identified various RD types from CXRs [1,14,15,19,20]. Especially, the VGG-based architecture [21] exploited transfer learning for detecting pneumonia by localizing the affected area in CXRs. While this approach achieved higher precision, the need for an individual network for detection and localization amplifies its complexity. Alternatively, [22] fused handcrafted features extracted from the segmented lung region using the Info-MGAN network for effective disease classification. However, training the MGAN network is quite challenging and its performance was extremely sensitive to the number of labeled samples and used parameters. Perhaps, the unreliable and less interpretable nature of the above DL models restrain their extension to real-time diagnosis. To gain interpretability [23] realized the Multi-task deep-end-to-end COMiT-Net to diagnose infections from CXRs followed by segmentation of lesion regions. Likewise, [24] used an explainable artificial intelligence (XAI) model that localized the area with the greatest likelihood of categorized images using a trained CNN model. However, the inclusion of numerous models significantly increases the model's complexity while decreasing its accuracy.

Alternatively, [25] utilized the task-specific adversarial network to increase the domain-invariant features and an uncertainty-aware ensembling model for predicting unseen data. The dilation varying CNN (CovXNets) [13] extracted features across various resolutions and performed localization using the integrated gradient-based discriminative model for distinguishing the abnormal regions. The confidence-aware anomaly detection (CAAD) model [26], predicted abnormal regions in CXR using low confidence and high anomaly detection scores. Likewise, cascaded rotation-invariant augmented hand features [11] coupled with DL in the ensemble framework effectively discriminated lesions based on the rank of probability score. Nevertheless, the integration of numerous DL networks into the aforementioned model makes it extremely complicated. Alternatively, the semi-supervised open set domain adversarial network (SODA) [27] effectively aligned data distributions at both the domain and subspace levels, and extracted features that were closely related to pathology location, achieving improved classification performance. Alternately, the memristive crossbar CNN array [28] classified the tunable Q-wavelet features acquired from decomposed CXRs for better storage, and quicker processing. Similarly, the anatomy-aware (AA) DL model [8], underlined the anatomical information for severity determination of pneumonia by learning crucial features at the disease level employing lung involvement scores. Although these DL variants provided superior accuracy, their black-box nature and high reliability on known targets question their extension to random real-time environments comprising unexpected outcomes. Also, realizing them on a simple hardware framework is practically less feasible thereby, demanding high-performing GPUs insisting on several weeks for model training. Also, an array of DL limitations refrain from their usage in medical imaging analysis including the lack of domain-specific knowledge and the position, density, and spread of fluid, and space varying significantly between cases, thereby, making it difficult to develop a one-size-fits-all model. In addition, their lack of transparency with interpretability is notoriously challenging when employed in medical diagnostics, where practitioners are uncertain of the arrived conclusion, which is a potential pitfall. Thereby, mandating well-defined and simple mathematical models in real-time diagnosis.

Analytical and machine learning models, on the other hand, are simple, easy to train, deploy, quick, and ensure reliability. Furthermore, ground-glass opacities derived from CXRs characterized textual information concerned with the parametric tuning of their patterns aiding accurate diagnosis [29] across diverse imaging modalities. The textual model [30], determined the RDs in CXR images using the Nearest Neighbour scheme using wavelet coefficients. The method performed moderate classification despite being simple. The fuzzy c means [31] clustered weighted pixels for pneumonia detection from frequency-transformed CXR images using the stationary wavelet transform. The pixel-wise computations incurred more time than their peers. To annotate lesions on CXR images [32], utilized a Multiple-Instance utilizing Support Vector Machine (miSVM) tuned by occurrence probabilities for accurate label prediction, rather struggled in lesion localization. This uncertainty was addressed in [32], [33] including miSVM in an active learning framework to localize lesions using expert relabeling. Instead [34] integrated multiscale statistical shape characteristics with various textural features of distinct lung boundaries using multiresolution approaches for boosting stability with accuracy while the model was less noise resilient. Rather, [4] ensured uniformity by augmenting the lung-segmented CXR images using affine transformations followed by a pair-wise comparison to determine the pneumonia likelihood using the EMD (Earth Mover Distance). Similarly [35] ranked the weighted entropies of multiple-criteria decision models for COVID-19 offering enhanced classification accuracy. The merger of several models considerably increased the process complexity, making it unsuitable for real-time applications.

The Random Forest (RF) variant of [36] selected crucial features based on the permutation score produced by the Boruta technique for the assessment of COVID-19 risk. To reduce the complexity, the RF classifier generated an associative tree that was unsuitable for imbalanced datasets. While [37], coupled Shannon entropy with fractal dimensions for the detection of pneumonia in CXR. However, the method was unable to localize the infected area and its severity level based on the magnitude of the fractal dimension. An ensemble of machine learning classifiers [6] quantified pneumonia severity based on the selection of integrated handcraft features using Principal Component Analysis (PCA) with Recursive Feature Elimination (RFE) techniques. Similarly, [38] blended fractal dimension, radiomics, and superpixel-based histon to diagnose pneumonia in CXRs registered higher accuracy while, its inability to distinguish the different infiltration patterns and the selection of the masking process offered less usage. The automatic diagnostic [39] detected infections based on the significant score (P-value<0.0001) by determining the left lower lung field's white pixels ratio on the segmented CXR. However, this hypothesis testing results were unreliable and completely relied on the sample size. To limit the influence of noisy labels, [40] weighted data distribution across multiple levels was performed followed by engagement of a variety of feature selection techniques based on the confidence scores that are then KNN classified as COVID-19 or not. Despite their usefulness, the aforementioned techniques demand complicated algorithms and are less effective than DL models when it comes to identifying lung infections. The pitfalls of both the black-box and white-box models necessitate the quest for suitable alternatives ensuring a trade-off between accuracy and complexity when extended to automatic real-time diagnosis offering reliability. As a result, a novel and computationally feasible automated model for the detection and classification of diverse RDs from CXRs using a pixel-wise approach is introduced by adopting the following strategies:

1. Primally the framework engages the novel Exponential Pixelating Integral (EPI) for the effective delineation of low-contrast regions with noise resilience
2. Feature representation deploying locally invariant fractal geometries namely the Mandelbrot and Julia sets for effective discerning of prominent textural features

3. Model classification using the ensembled Multivariate Adaptive Regression Splines (MARS) ensuring robustness with generalization

Detailed evaluations of the developed system on 100,000 CXRs obtained from various publicly available large datasets such as Kaggle, RSNA, and NIH are conducted. The observations reveal that EPI merged with fractal geometry improved classification performance, demonstrating its potential as a valuable means for automated analysis of CXRs in a clinical setting. Accordingly, the presented content is compartmentalized as follows: EPI formulation and application of fractal geometry for extracting the essential features coupled with the MARS classifier is dealt with in Section 2. An exhaustive evaluation of the diagnostic model is done in Section 3 followed by the simplicity analysis in Section 4. The conclusion in Section 5 outlines the proposal's intention with its suitability for real-time diagnostics.

## 2. Methodology

Diagnosing and appropriately treating RDs based on CXRs poses a challenge due to the inherent similarities among these disorders and the superimposed nature of anatomical structures. Moreover, another key challenge in the analysis of CXR images is the small differential intensity values in grayscale images, which can limit the focus on smaller scales. To overcome this limitation, a new image transformation technique called Exponential Pixelating Integral (EPI) is proposed. EPI computes the exponential moving average over a set of pixels bound by an overlapping local 3x3 kernel for contrast enhancement by sidelining the subtle intensity variations observed between adjacent pixels, hence facilitating the extraction of more meaningful features for classification. Subsequently, the resulting EPI image is polar transformed to facilitate a convenient means of representing and manipulating complex numbers and symmetrical forms in comparison to alternative coordinate systems. The polar transformed image is inputted to the Mandelbrot and Julia sets to produce a fractal representation covering complex shapes thereby systematically capturing object roughness which is extremely essential in Feature characterization. Especially, fractal geometry holds significant utility in the realm of medical diagnosis, particularly in lesion detection. The detection of malignant cells is facilitated by the contrasting growth patterns exhibited by healthy human blood vessel cells, which normally adhere to an organized fractal arrangement, in contrast to the aberrant growth exhibited by malignant cells. The utilization of fractal analysis facilitates the differentiation between normal tissue structures and indicators of potential abnormalities. Later, a wide range of statistical characteristics are extracted from the EPIFF that are then fed to the ensemble Multivariate Adaptive Regression Splines (MARS) for pair-wise comparison between diverse RDs favoring robustness with generalization. The aforementioned process is depicted in Fig. 1 including in-depth discussions of the aforementioned objective, dealt with in the appropriate subsections.

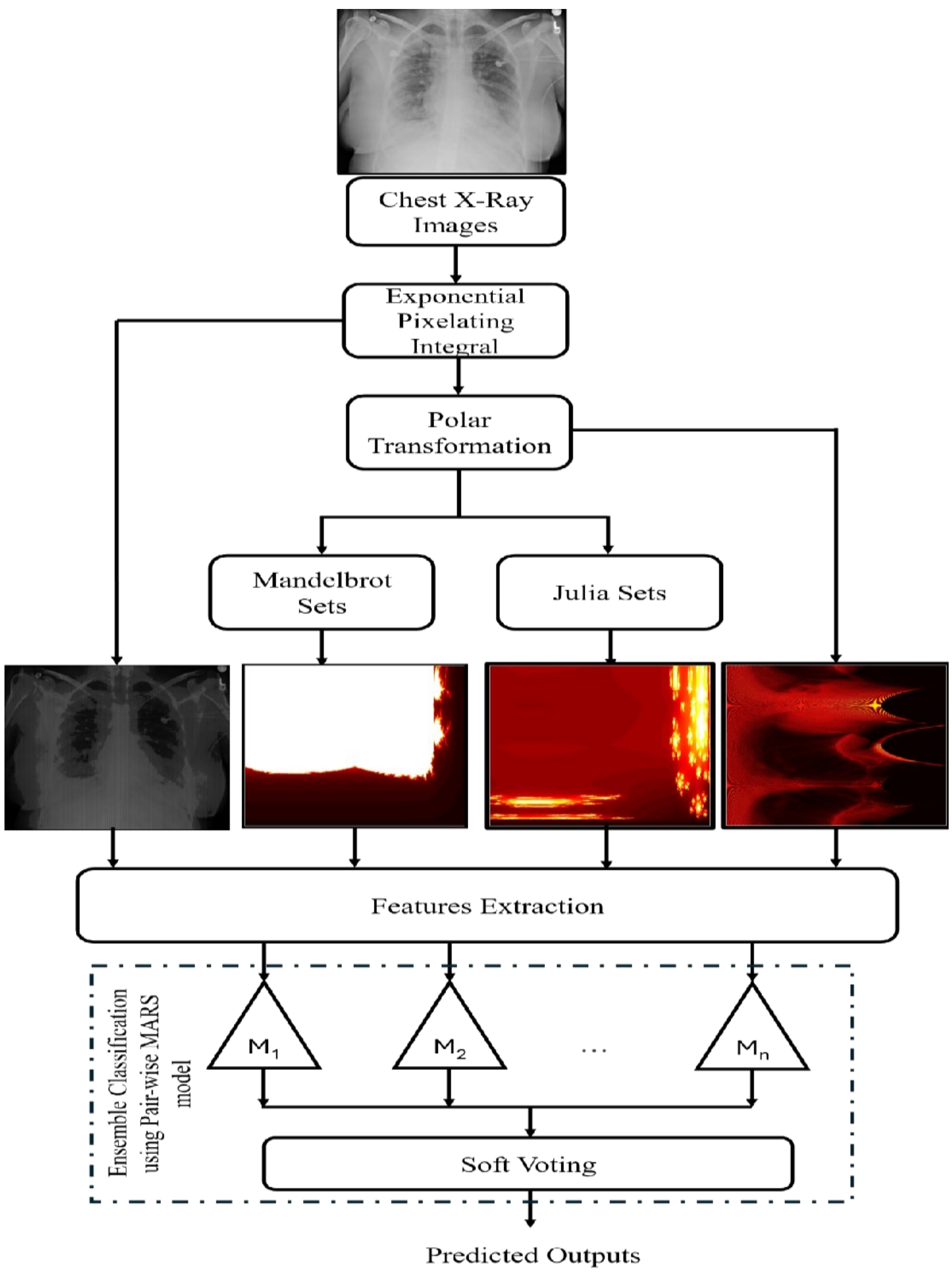

**Fig.1 Process diagram of the proposed EPIFF-MARS diagnostic framework**

### 2.1 Exponential Pixelating Integral (EPI) model

Detection of lesions on a CXR image becomes challenging when there is relatively low contrast between the lesion and the surrounding tissue. To tackle this issue, several methodologies have employed contrast enhancement strategies aimed at increasing either the global or local contrast, to improve the the visual clarity of the image [9]. Though contrast-enhancing techniques decrease the occurrence of misinterpretation, they add noise and introduce distortions at the finer level. As a result, developing a simple solution that successfully addresses this contrast issue without increasing the noise effect remains a challenge. To address these problems, an innovative Exponential Pixelating Integral model is introduced for successfully highlighting the overall intensity with noise suppression. The modeled EPI smoothens the image with simultaneous noise suppression thereby reducing its visibility. Owing to the overlapping localized EPI nature, the resultant image will undergo scaling and standardization thereby enhancing the lower-intensity difference regions in a wider sense. This highlights the lesions, which are efficiently utilized to reference pixel-related fractal sets.

Accordingly, to get the EPI image the process commences with resizing the input image $I$ into $m \times n$ dimensions ($I \epsilon R^{m \times n}$) to maintain uniformity between extracted features of CXRs in the dataset. The resultant image is then parted into $r \times r$ non-overlapping subregions to extract the localized features.

Let $K$ be the kernel that represents subregions with the size $(r = 3)$ of the given input image as defined in Eq.1

$$K_{r \times r} = [p_{i-1,j-1}\ p_{i-1,j}\ p_{i-1,j+1}\ p_{i,j-1}\ p_{i,j}\ p_{i,j+1}\ p_{i+1,j-1}\ p_{i+1,j}\ p_{i+1,j+1}] \quad (1)$$

Where $p_{ij}$ − pixel value of the $K_{r \times r}$ at image $I \epsilon R^{m \times n}$. Let $p_{i-1,j-1}$ and $p_{i-1,j}$ be the two consecutive pixels, then the intensity difference between them is denoted as $\Delta p$ and is defined in Eq.2

$$\Delta p = |p_{i-1,j-1} - p_{i-1,j}| \quad (2)$$

Herein, the instant neighbor is written as $p_{i-1,j} = p_{i-1,j-1} \pm \Delta p$. To determine the local variance from the extracted kernel elements, initially, mean $(\mu)$ is calculated between $p_{i-1,j-1}$ and $p_{i-1,j}$ using Eq. 3.

$$\mu = \frac{p_{i-1,j-1} + p_{i-1,j}}{2} \quad (3)$$

Substituting $p_{i-1,j}$ in Eq. 3 results in the approximation of $\mu$ as expressed in Eq.4

$$\mu = p_{i-1,j-1} \pm \frac{\Delta p}{2} \quad (4)$$

Consequently, the variance between two consecutive pixels $p_{i-1,j-1}$ and $p_{i-1,j}$ is evaluated using Eq.5

$$\sigma^2 = \frac{(p_{i-1,j-1} - \mu)^2 + (p_{i-1,j} - \mu)^2}{2} \quad (5)$$

Equation 6 is obtained by substituting Eq.4 in Eq.5

$$\sigma^2 = \frac{\left(p_{i-1,j-1} - p_{i-1,j-1} \mp \frac{\Delta p}{2}\right)^2 + \left(p_{i-1,j-1} \pm \Delta p - p_{i-1,j-1} \mp \frac{\Delta p}{2}\right)^2}{2} \quad (6)$$

Simplifying Eq.6 produces the Variance of $p_{i-1,j-1}$ and $p_{i-1,j}$ in Eq.7

$$\sigma^2 = \frac{\left(\left(\frac{\mp \Delta p}{2}\right)^2 + \left(\pm \Delta p \mp \frac{\Delta p}{2}\right)^2\right)}{2} = \frac{\Delta p^2}{4} \quad (7)$$

Equation 7 represents the variance between the two consequent pixels of the normal kernel function over an image. The acquired variance is insufficient to capture fine subtle texture details in low-contrast CXR images which are crucial for diagnostic purposes. Hence, the process introduces the Exponential Pixelating Integral Transform (EPI) for enhancing the granular image details. To achieve that the process transforms the original image into the EPI equivalent by applying the kernel $\hat{k}$ which evaluates the exponential function of the cumulative sum of neighboring pixel intensities, allowing a greater degree of emphasis on minute-scale features and is described in Eq. 8

$$\hat{k}(i,j) = [\, e^{p_{i-1,j-1}}\ e^{p_{i-1,j-1}+p_{i-1,j}}\ e^{p_{i-1,j-1}+p_{i-1,j}+p_{i-1,j+1}}\ e^{p_{i,j-1}}\ e^{p_{i,j-1}+p_{i,j}}\ e^{p_{i,j-1}+p_{i,j}+p_{i,j+1}}\ e^{p_{i+1,j-1}}\ e^{p_{i+1,j-1}+p_{i+1,j}}\ e^{p_{i+1,j-1}+p_{i+1,j}+p_{i+1,j+1}} \,] \quad (8)$$

In Eq. 8, the cumulative sum captures the accumulation of intensity changes within a local area creating a more prominent peak in the edge region. In addition, the exponential function amplifies intensity variations in input images, causing small changes to result in significant variations in the transformed images. Thus, the exponential of the cumulative sum effectively visualizes the localized intensity variations. Later, the mean $(\mu_e)$ of EPI transformed image is determined between consecutive pixels $e^{p_{i-1,j-1}}$ and $e^{p_{i-1,j-1}+p_{i-1,j}}$ present in the kernel is evaluated using Eq.9

$$\mu_e = \frac{e^{p_{i-1,j-1}} + e^{p_{i-1,j-1}+p_{i-1,j}}}{2} \quad (9)$$

Upon substituting $p_{i-1,j} = p_{i-1,j-1} \pm \Delta p$ in Eq.9 yields Eq.10

$$\mu_e = \frac{e^{p_{i-1,j-1}} + e^{2p_{i-1,j-1} \pm \Delta p}}{2} \quad (10)$$

Finally, the mean determined in Eq.11 is obtained by simplifying Eq.10

$$\mu_e = = \frac{e^{p_{i-1,j-1}} + \left(e^{p_{i-1,j-1}}\right)^2 e^{\pm\Delta p}}{2} \tag{11}$$

The mean in Eq.11 is subtracted from the pixels $e^{p_{i-1,j-1}}$ and $e^{p_{i-1,j-1}+p_{i-1,j}}$ and squared to produce the variance as in Eq.12

$$\sigma_e{}^2 = \frac{\left(e^{p_{i-1,j-1}} - \mu_e\right)^2 + \left(e^{p_{i-1,j-1}+p_{i-1,j}} - \mu_e\right)^2}{2} \tag{12}$$

Substituting Eq. 11 in Eq. 12 results in Eq.13

$$\sigma^2 = \frac{\left(e^{p_{i-1,j-1}} - \frac{e^{p_{i-1,j-1}}}{2} - \frac{\left(e^{p_{i-1,j-1}}\right)^2 . e^{\pm\Delta p}}{2}\right)^2 + \left(e^{p_{i-1,j-1}+p_{i-1,j-1}\pm\Delta p} - \frac{e^{p_{i-1,j-1}}}{2} - \frac{\left(e^{p_{i-1,j-1}}\right)^2 . e^{\pm\Delta p}}{2}\right)^2}{2} \tag{13}$$

Eq.13 is simplified to yield Eq.14

$$\sigma^2 = \frac{\left(e^{p_{i-1,j-1}}\right)^2}{4} \cdot \left(e^{p_{i-1,j-1}} . e^{\pm\Delta p} - 1\right)^2 \tag{14}$$

The $e^{\pm\Delta p}$ term in Eq.14 enhances the variance even for a small intensity difference, thereby, intensifying EPI to focus more on minuscule features. To further contrast stretching, the process commences with the normalization of individual pixels in the EPI-transformed kernel $\hat{k}(i,j)$ given in Eq. 8 using Eq.15

$$\hat{g}(i,j) = e^{\sum_{k=1}^{j} p_{i,k}} \tag{15}$$

To remove the offset caused by illumination across the image and normalize the spread of intensity across the kernel, each pixel value is normalized by its row-wise mean and row-wise variance of the kernel $(\gamma)$ as in Eq. 16

$$\underline{p}_{i,j} = \gamma \times \hat{g}(i,j) \ \ where \ \ \gamma = \left(\frac{p_{i,j}}{(\varepsilon+\mu_i)*(\varepsilon+\sigma_i^2)}\right) \tag{16}$$

Where $\varepsilon$ a small constant is assumed to be 0.001 and is introduced to avoid division by zero thereby, ensuring numerical stability. This normalization removes the outliers and improves the classification accuracy by increasing the discrimination between samples. After normalizing, to mitigate variations in illumination and highlight subtle features for accurate diagnostic interpretation, the final transformed intensity values are enhanced using contrast stretching as expressed in Eq.17

$$\hat{p}_{i,j} = \{\underline{p}_{i,j} + \sigma \quad if \ \underline{p}_{i,j} \geq \mu \ \ \underline{p}_{i,j} - \sigma \quad if \ \underline{p}_{i,j} < \mu \tag{17}$$

The normalized exponential cumulative sum followed by contrast stretching of each pixel in the kernel highlights the possible lesions and regions of interest, resulting in enhanced discrimination. Finally, the kernel function yielded upon applying the aforesaid modifications is given in Eq.18

$$K = [\hat{p}(i-1,j-1)\ \hat{p}(i-1,j)\ \hat{p}(i-1,j+1)\ \hat{p}(i,j-1)\ \hat{p}(i,j)\ \hat{p}(i,j+1)\ \hat{p}(i+1,j-1)\ \hat{p}(i+1,j)\ \hat{p}(i+1,j+1)\ ] \tag{18}$$

A snapshot of the EPI-transformed images is shown in Fig. 2 wherein effective lesion distinctions are noticed by emphasizing the relevant regions.

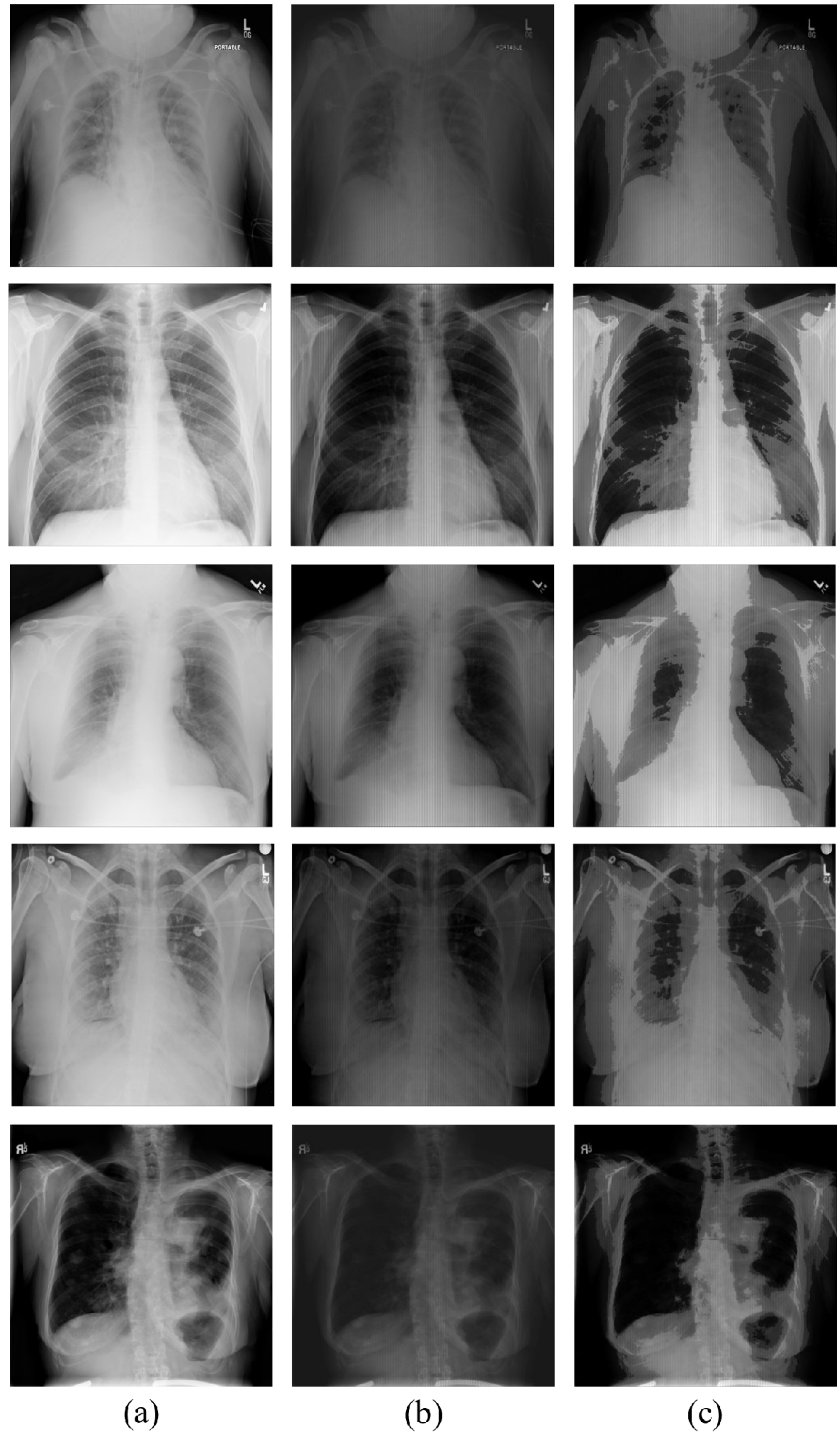

**Fig 2. Resultant EPI transformation on the CXR images. (a) The original X-ray images, (b) Normalized EPI transformed images, (c) Contrast Stretched images.**

From Fig. 2 it is understood that the EPI transform effectively highlights the subtle variations in pixel intensity, making it easier to identify and analyze specific features within an image. Upon attainment of the EPI transformed image additionally two more feature matrices are extracted for distinguishing the structural abnormalities by considering the local affinity with invariance towards affine variations and is elaborated below.

### 2.2 Fractal Representation

Fractal geometry is a fascinating subject with substantial applications in a variety of medical image-processing techniques for analyzing the sophisticated and diverse features present in a wide range of biological systems[41]. A prominent benefit of these sets is in their self-similarity, whereby the identical fundamental pattern is replicated at varying scales. This characteristic makes them extremely suitable for representing intricate and non-uniform patterns posing challenges. Accordingly, the Mandelbrot and Julia sets are constructed in this approach straightforwardly using analytical models[42]. This facilitates better interpretation of complex structures, leading to more accurate diagnosis as the regular fractal growth pattern exhibited commonly in human organ tissues aids in the detection of malignant cells that grow erratically. Therefore, engaging fractal analysis eases the structural distinction between normal and abnormal structures, enhancing diagnosis accuracy[43]. To determine the fractal sets, the EPI image is initially converted into polar coordinates owing to its simplicity in representation and manipulation of symmetrical shapes. In the polar coordinate system, each point on a plane is characterized by its distance from a reference point $(r)$ and its angular displacement $(\theta)$ from a reference direction. The utilization of the $(r)$and $(\theta)$ parameters enhance the representation of texture variation by establishing a direct correlation between image features and the directionality of the texture [44], and one such outcome is demonstrated in Fig. 3 (a).

#### 2.2.1 Mandelbrot Image Representation

The choice of Mandelbrot and Julia sets is attributed to identifying similar data patterns and significant attributes within images. In addition, their simplicity in generation yet powerful in representation, make them a popular tool in the field of fractal geometry. Mathematically, a Mandelbrot set represents a set of complex numbers as in Eq.19

$$f(z) = z^2 + c \tag{19}$$

Where $z -$ is the independent variable and initially assumed to be zero, $c$ is a complex number. The function is iterated a fixed number of times, and if the absolute value of the result remains less than a certain threshold, the point $c$ is considered to be in the set.

To generate a mandelbrot set, the process first determines the standard deviation of the polar image along the rows $(\sigma_i)$ and columns $\left(\sigma_j\right)$ that are then composed as the real and imaginary components of the complex additive [45] as defined in Eq. 20

$$c = \sigma_i + i\sigma_j \tag{20}$$

Where, $\sigma_i$ *and* $\sigma_j -$ bounded in the range $i \in [0, m]$ *and* $j \in [0, n]$, $m, n$-be the size of the polar image.

Finally, the Mandelbrot set is derived through the iterative process by substituting Eq.20 into Eq.19, resulting in Eq. 21

$$f_c(z) = |z_{n+1}| = z_n^2 + \left(\sigma_i + i\sigma_j\right) \tag{21}$$

$|z_n|$ undergoes iterations $(t)$ until surpassing a divergence threshold and is denoted as $|z_{n+1}| \leq 2$. If the absolute value of the result remains less than 2, then only the point $c$ is considered to be in the set.

Accordingly, the mandelbrot set is constructed using Eq. 22

$$M = \{c \in C : f^0(c) \in i \,;\, f^t(c) \to \infty \; as \; t \to \infty\} \tag{22}$$

where $C -$ set of complex numbers; $f^t(c) -$ iterates of the function. This configuration connects the Real and Imaginary axes, with the complex plane represented by black space. There are two sections to the structure shown in Fig.3(b) the main bulb with the main cardioid. These elements tend to compress, expand, rotate, or translate due to the nature of the standard deviation along both axes. The variance-dependent Mandelbrot set generates definitive patterns subject to the class. The boundary, and area of spread clearly define the Region of Interest for feature extraction in the Mandelbrot set represented in Fig. 3(b).

### 2.2.2 Julia Set Image Representation

Technically, the Julia Set structure has strands of lines and infinitely many pieces. Such a structure is termed Fatou Dust [46]. In this set, the complex additive is a constant while the z is the point of interest. This set exhibits similarities to the Mandelbrot set, with the key distinction lying in their respective plotting methodologies. While the Mandelbrot set encompasses the entire range of possible values for c, the Julia set focuses on the visualization of a specific value of complex additive $(c)$ and is defined in Eq.23

$$f(z) = z^2 + c \tag{23}$$

The complex additive $c$ is constant and bounded between [0,1]. While, the $z$ value is dependent on the standard deviation of the polar transformed image, across row and column as the real and imaginary parts respectively as described in Eq.24

$$z_n = \sigma_i + i\sigma_j \tag{24}$$

Finally, the Julia Set is derived through the substitution of Eq.24 into Eq.23, resulting in Eq.25

$$f_c(z) = |z_{n+1}| = z_n^2 + c \tag{25}$$

The initial value of $z$ is assumed zero (denoted in complex form as $0 + 0i$) to avoid divergence to infinity. Similar to the mandelbrot set, the process iterates until it reaches the threshold value of $|z_{n+1}| \leq 2$.

Finally, as per the basin of attraction to infinity function, the Julia Set is constructed using Eq. 26

$$J = \{z \in C : f^n(z) \to \infty \; as \; n \to \infty \} \tag{26}$$

where $C$ is the set of complex numbers and $f^n(z)$ are the iterates of the function.

The Julia set generates multiple clusters with varying intensities and definitive boundaries. Feature extraction concerning the cluster intensities and their variation are used to analyze the mapped source image depicted in Fig. 3(c). In comparison with other image transforms and representations, these fractal sets extract multiple features and are highly tunable enabling effective representation of minuscule image details.

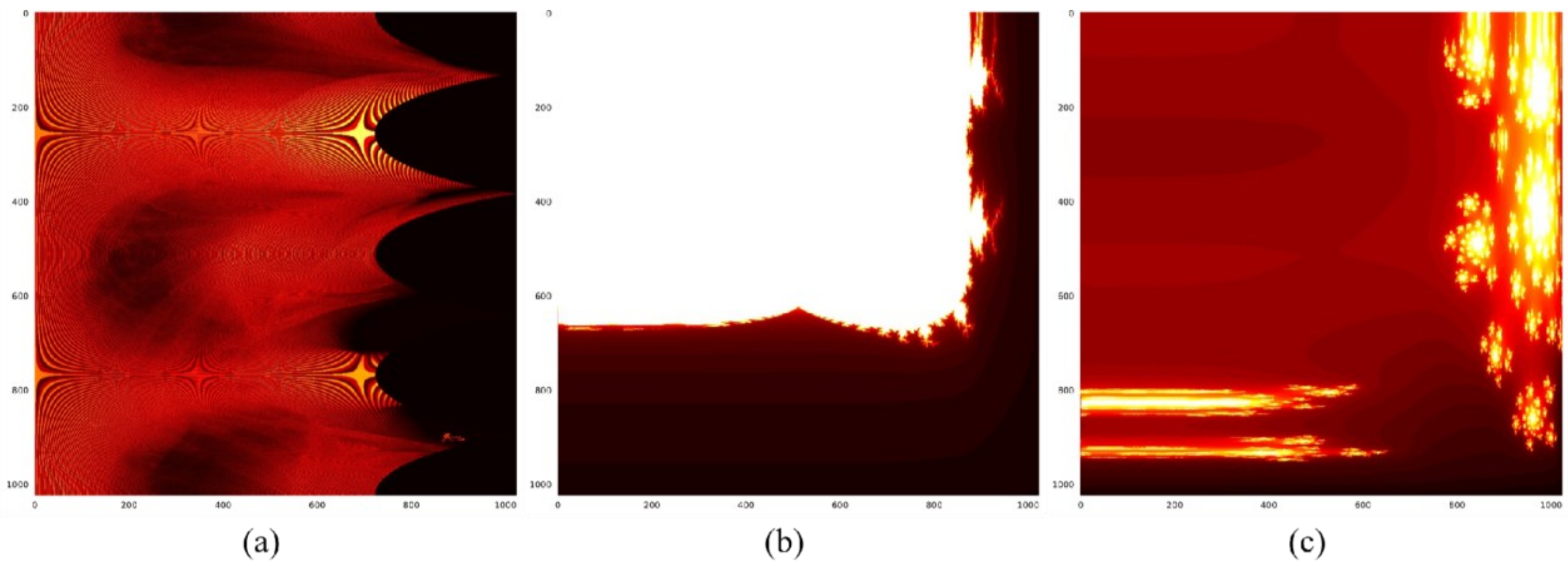

**Fig.3 Image Representation (a) Polar Image (b) Mandelbrot set, and (c) Julia set Representation**

Consequently, to precisely classify various RDs, EPIFF extracted the multiple statistical features from the aforementioned image representation namely, the Polar transformed image, and both the Julia & Mandelbrot Image representation which are detailed with notation presented in Table 1. These features attributed to valuable information for different aspects of image analysis, including structure, intensity distribution, and statistical properties.

**Table 1 Features extracted from Polar Transformation, Mandelbrot set, and Julia set Representation.**

| Features | Symbols | | | Description |
|---|---|---|---|---|
| | Polar | Mandelbrot | Julia | |
| Binarized Sum | $p_{bin}$ | $m_b$ | $j_b$ | Provides a measure of the total intensity in a binary image |
| Centroid for x-axis | $pc_x$ | $mc_x$ | $jc_x$ | Computes the Centroid:x of the corresponding image contour for localization |
| Centroid for y-axis | $pc_y$ | $mc_y$ | $jc_y$ | Computes the Centroid:y of the corresponding image contour for localization |
| Euler Number | $p_{eul}$ | $m_{eul}$ | $j_{eul}$ | Calculate the Euler number of the corresponding image contour for shape characterization |
| Mean | $p_{me}$ | $m_{me}$ | $j_{me}$ | Average intensity of pixels in an image |
| Variance | $p_{var}$ | $m_{var}$ | $j_{var}$ | Quantifies the spread or dispersion of pixel intensities to provide textural information |
| Dominant Pixel Frequency | $p_{df}$ | $m_{df}$ | $j_{df}$ | Highest occurring pixel intensity to diagnose predominant regions |
| Dominant Pixel | $p_{dp}$ | $m_{dp}$ | $j_{dp}$ | Highest occurring pixel intensity to detect prominent textures |
| Median | $p_{med}$ | $m_{med}$ | $j_{med}$ | The median of the pixel intensities |
| Median Frequency | $p_{mf}$ | $m_{mf}$ | $j_{mf}$ | Frequency of Median of pixel intensities to measure the distribution of median pixels |
| Kurtosis | $p_{kt}$ | $m_{kt}$ | $j_{kt}$ | Quantify the peakedness of distribution to characterize the shape of intensities |
| Skew | $p_{sk}$ | $m_{sk}$ | $j_{sk}$ | Measure the asymmetry of the distribution reveals the distribution of pixel intensities |
| Standard Error of the Mean | $p_{sem}$ | $m_{sem}$ | $j_{sem}$ | SEM measures the average deviation of sample means from the true population mean. |
| Coefficient of Variation | $p_{cv}$ | $m_{cv}$ | $j_{cv}$ | Computes the coefficient of variation, which is the ratio of the standard deviation to the mean of the data. |

Further, the process extracted a few more essential features that are used to estimate the characteristics of the proposed EPI-transformed images in comparison with normal CXRs. These features are valuable for capturing subtle yet significant alterations in pixel values across the images aiding in diagnosing and understanding the disease's effect which are detailed in Table 2.

**Table 2 Features extracted from EPI transformed images**

| Features | Symbol | Description |
|---|---|---|

| Peak Signal-to-Noise Ratio | $psnr$ | Quantifies the disease fidelity, while higher values indicate that the disease-induced changes are minimal, ensuring that relevant anatomical features are preserved. |
|---|---|---|
| Spearman Correlation | $sp$ | Measures the degree of association. Also, identify the consistent changes in pixel values across the images due to disease-induced alterations. If a high correlation is observed, it suggests that certain features are consistently affected by the disease. |
| Mean squared error | $mse$ | Estimate the extent of disease-related changes, aiding in identifying regions with significant alterations. Larger MSE values point to areas of pronounced differences, which could indicate disease presence with severity. |
| Structural Similarity | $ssim$ | Quantifies image quality degradation. SSIM considers luminance, contrast, and structure, aligning with radiologists' interpretation of images. A lower SSIM suggests that disease-related structural changes impacting image similarity |
| Normalized MSE | $n_{mse}$ | Computes MSE at different scales to enhance the interpretability |
| Information of Case \| Normal | $icn$ | Evaluates the predictive ability of the model when the case information is available |
| Information of Normal \| Case | $inc$ | Computes mutual information of normal, given a Case, offering insights into the significance of the normal class. |
| Information Score | $i_{sc}$ | Clustered mutual information between two images. It's important to assess the degree to which disease alters relevant image features. A high Information Score could highlight regions of high diagnostic relevance. |
| Euclidean Distance | $eu$ | Computes Euclidean array between two arrays to measure the similarity |
| Energy Distance | $e_{dist}$ | Quantify the dissimilarity between images based on pixel values and distributional differences. They identify the global shifts caused by diseases. Higher distances indicate widespread alterations, offering insights into disease progression and impact. |

Overall 52 essential features listed in Tables 1-2 are extracted from the EPI, Polar, and dual fractals accumulated to yield the Feature Vector (FV) of a given CXR. Though the feature length seems small, which is extracted from the EPI, polar transformed images followed by Mandelbrot and Julia set representation. EPI transformation enhances the low-contrast regions while polar conversion significantly highlights the curved surfaces. These transformations are further exploited by the inherent self-similarity nature of fractals to capture the structural distinction between normal and abnormal regions, resulting in compact and highly discriminating features that escalate the diagnostic accuracy with reduced complexity. Further, this concise set of features facilitates rapid prediction which is extremely crucial for real-time applications. Likewise, the FVs' of dataset constituents are extracted and formulated into a feature database that is to be classified by considering the diverse criteria concerned with RDs. To cover the wider aspects of various RDs a multivariate regression model adapting to intricate data changes is adopted and dealt with in the below section.

**2.3 Classification using MARS Ensemble Learning**

Generalized Linear and Additive Models assume that coefficients associated with predictor variables remain constant across all predictor values. However, MARS discards assumptions and is non-parametric[47,48]. It uses a combination of basis functions, such as polynomials or splines, to approximate the relationship between the input and output variables. This makes it well suited for complex and highly varied data such as medical images, as it can adapt to the underlying patterns in the data without making strong assumptions about the shape of the relationship. Moreover, the MARS offers more interpretability than deep learning models by relatively weighing the different input features aiding prediction. Deep learning models that represent black boxes naturally make it difficult to understand the decided predictions and are more complex.

MARS model is mathematically defined in Eq. 27

$$y = b_0 + b_1 x_1 + b_2 x_2 + \cdots + b_n x_n + \sum_{i=0}^{n} \alpha_i h_i(z_i) \quad (27)$$

where $y$ −output variable, $b_i$ −coefficient of linear terms, $x_i$ −input variable, $\alpha_i$ −coefficient of nonlinear terms, $h_i(z_i)$ − Basis functions.

The classification model is built in two phases by the MARS. In the forward phase, it systematically incorporates Basis Functions (BF) into its structure to minimize the mean squared residual error. This entails evaluating each variable in the dataset as a potential basis function until a negligible change in the residual error is observed. The BF takes the largest value from one of two options: 0 or the outcome of the environmental variable – knot value and is defined in Eq.(28)

$$BF = max\ (0, env\ var - \ knot) \quad (28)$$

Where, $env\ var$ −individual feature elements in FVs. By raising the value of the basis function, MARS produces a persistent estimate of the target function to an arbitrary order of derivatives. A distinct linear regression with its slope is constructed for each group while Knots link them automatically in an optimal manner. In contrast, the backward phase involves pruning non-participating basis functions that do not contribute to improving model performance. By eliminating irrelevant features through backward elimination, MARS reduces model complexity. Unlike traditional models that might use a generic set of features across all categories, MARS analyzes the nature of the input data and identifies the most critical features specific to each category. This streamlined model requires less computational power to train and employ, making it more efficient for handling large datasets. While feature selection improves scalability, it's crucial to consider the impact on accuracy. MARS addresses this through a rigorous pruning process wherein features having minimal impact on its performance are discarded based on the Generalized Cross-Validation and preserve the most essential features for accurate predictions.

Accordingly, to classify the diverse RDs, a pair-wise comparison of each disease ($D_i$) against every other namely Edema, COVID-19, Effusion, Emphysema, and Pneumonia is performed. As the diseases were categorized into five, hence, the class-wise comparisons led to 5 thereby resulting in 25 (5x5) possible comparisons. Accordingly, 25 MARS models are trained on 80% of the data with each model specialized in differentiating a specific disease pair and identifying key features in the form of basis function and its coefficients during training. The remaining 20% of the data validates these models. When classifying a new data point, the ensemble learning mechanism comes into play wherein the remaining four models relevant to the disease (one for each comparison with other diseases) make predictions. These predictions are then combined by soft voting technique to determine the most likely disease. This approach leverages the strengths of multiple, specialized models, focusing on specific disease differentiation.

Accordingly, EPIFF-MARS formulates basis functions ($A_{ij} \in R^{F\times k}$) along with its coefficient matrix ($C_{ij} \in R^{F\times k}$) for each category separately. where $F$ −number of participant features $F$; $k$ −total number of categories; $i, j$ − denotes pairwise comparison between disease categories bounded between $[0, k - 1]$;

For instance, to classify Edema, the model initiates training by tuning the basis function and corresponding coefficient matrix from the EPIFF features related to Edema with the remaining classes as, Edema vs. edema ($A_{0,0}$, $C_{0,0}$); Edema Vs COVID-19 ($A_{0,1}$, $C_{0,1}$); Edema vs. effusion ($A_{0,2}$, $C_{0,2}$); Edema vs. emphysema ($A_{0,3}$, $C_{0,3}$); and Edema Vs Pneumonia ($A_{0,4}$, $C_{0,4}$) resulting in five distinct $A_{ij}$ with $C_{ij}$ to build the MARS model related to Edema assuming $i = 0$ as in Eq. (29)

$$D_0(Edema) = \left[C_{0,0}^T \cdot A_{0,0} + C_{0,1}^T \cdot A_{0,1} + C_{0,2}^T \cdot A_{0,2} + C_{0,3}^T \cdot A_{0,3} + C_{0,4}^T \cdot A_{0,4}\right] \quad (29)$$

Eq. 29 is generalized as in Eq. 30 for training the MARS model.

$$D_i = \sum_{j=0}^{k-1} \left(C_{ij}\right)^T \cdot A_{ij}\big|_i \quad (30)$$

$D_i$ −represents Disease belonging to $i^{th}$ category.

Similarly, upon adopting the aforesaid approach the overall MARS classification model is constructed in Eq. 31

$$[D_0(Edema)\ D_1(Covid-19)\ D_2(Effusion)\ D_3(Emphysema)\ D_4(Pneumonia)\ ] = \left[C_{0,0}^T \cdot A_{0,0} + C_{0,1}^T \cdot A_{0,1} + C_{0,2}^T \cdot A_{0,2} + C_{0,3}^T \cdot A_{0,3} + C_{0,4}^T \cdot A_{0,4}\ C_{1,0}^T \cdot A_{1,0} + C_{1,1}^T \cdot A_{1,1} + C_{1,2}^T \cdot A_{1,2} + C_{1,3}^T \cdot A_{1,3} + C_{1,4}^T \cdot A_{1,4}\ C_{2,0}^T \cdot A_{2,0} + C_{2,1}^T \cdot A_{2,1} + C_{2,2}^T \cdot A_{2,2} + C_{2,3}^T \cdot A_{2,3} + C_{2,4}^T \cdot A_{2,4}\ C_{3,0}^T \cdot A_{3,0} + C_{3,1}^T \cdot A_{3,1} + C_{3,2}^T \cdot A_{3,2} + C_{3,3}^T \cdot A_{3,3} + C_{3,4}^T \cdot A_{3,4}\ C_{4,0}^T \cdot A_{4,0} + C_{4,1}^T \cdot A_{4,1} + C_{4,2}^T \cdot A_{4,2} + C_{4,3}^T \cdot A_{4,3} + C_{4,4}^T \cdot A_{4,4}\ \right] \quad (31)$$

This model is mathematically shortened in Eq. 32

$$[D_0\ D_1\ D_2\ D_3\ D_4\ ] = \left[\sum_{j=0}^{k-1} \left(C_{ij}\right)^T \cdot A_{ij}\big|_{i=0}\ \sum_{j=0}^{k-1} \left(C_{ij}\right)^T \cdot A_{ij}\big|_{i=1}\ \sum_{j=0}^{k-1} \left(C_{ij}\right)^T \cdot A_{ij}\big|_{i=2}\ \sum_{j=0}^{k-1} \left(C_{ij}\right)^T \cdot A_{ij}\big|_{i=3}\ \sum_{j=0}^{k-1} \left(C_{ij}\right)^T \cdot A_{ij}\big|_{i=4}\right] \quad (32)$$

Based on the aforesaid construction, the pair-wise comparison of the MARS models for overall disease prediction with probability score is presented in Table 3.

**Table 3. Pairwise MARS models for overall disease prediction**

| Disease | Edema | Covid | Effusion | Emphysema | Pneumonia |
|---|---|---|---|---|---|
| **Edema** $\{> 0.5\ \ if\ G_T = Edema$ $< 0.5\ \ otherwise$ | 1 | $C_{0,1}^T \cdot A_{0,1}$ | $C_{0,2}^T \cdot A_{0,2}$ | $C_{0,3}^T \cdot A_{0,3}$ | $C_{0,4}^T \cdot A_{0,4}$ |
| **COVID** $\{> 0.5\ \ if\ G_T = Covid$ $< 0.5\ \ otherwise$ | $C_{1,0}^T \cdot A_{1,0}$ | 1 | $C_{1,2}^T \cdot A_{1,2}$ | $C_{1,3}^T \cdot A_{1,3}$ | $C_{1,4}^T \cdot A_{1,4}$ |
| **Effusion** $\{> 0.5\ \ if\ G_T = Effusion$ $< 0.5\ \ otherwise$ | $C_{2,0}^T \cdot A_{2,0}$ | $C_{2,1}^T \cdot A_{2,1}$ | 1 | $C_{2,3}^T \cdot A_{2,3}$ | $C_{2,4}^T \cdot A_{2,4}$ |
| **Emphysema** $\{> 0.5\ \ if\ G_T = Emphysema$ $< 0.5\ \ otherwise$ | $C_{3,0}^T \cdot A_{3,0}$ | $C_{3,1}^T \cdot A_{3,1}$ | $C_{3,2}^T \cdot A_{3,2}$ | 1 | $C_{3,4}^T \cdot A_{3,4}$ |
| **Pneumonia** | $C_{4,0}^T \cdot A_{4,0}$ | $C_{4,1}^T \cdot A_{4,1}$ | $C_{4,2}^T \cdot A_{4,2}$ | $C_{4,3}^T \cdot A_{4,3}$ | 1 |

$$\begin{cases} > 0.5 & if\ G_T = Pneumonia \\ < 0.5 & otherwise \end{cases}$$

Finally, the positive probability of a disease $d_i$ is determined by engaging the SoftMax function as defined in Eq.(33)

$$P(D_i) = \frac{e^{D_i}}{\sum_{j=0}^{k-1} e^{D_j}} \tag{33}$$

$P(D_i)$ produces a set of positive probabilities for each class associated with a given image by considering the maximum likelihood of occurrence to successfully predict the various RDs.

## 3. Performance Analysis

### 3.1 Datasets Description and Evaluation Metrics

To investigate the classification efficiency of the introduced EPIFF-MARS model, rigorous investigations were performed on Chest X-ray 14 and COVID-19 datasets. The Chest X-ray 14 dataset [49], is composed of 112,120 frontal-view CXRs acquired from 32,717 patients of the National Institutes of Health Clinical Center (NIH). The images are 8-bit gray-scale PNG files, sized 1024×1024 pixels, with an average of 3-4 images per patient due to follow-up scans. These images were initially associated with 8 thoracic pathologies and were later expanded to include 14 chest diseases[50]. The dataset includes metadata detailing pathologies, follow-up images, patient demographics, and radiography views. Among the 112,120 chest X-rays, 60,412 show no pathology, while 51,708 depict one or more pathologies. This dataset is valuable for chest radiography research and medical applications.

To demonstrate the EPIFF-MARS effectiveness in detecting COVID-19 abnormalities from CXRs, the experiment was conducted on the publically available COVID-19 Radiography dataset[51]. This dataset consists of 21,165 CXRs and their corresponding masks. Amongst them, 3,616 were imaged from COVID-19 patients with 6,012 showing non-COVID lung infections, 1,345 associated with pneumonia, and the remaining 10,912 were normal lung X-rays. The images are in PNG format with a dimension of 256 × 256 pixels. This dataset was created by the international collaboration of researchers from Qatar University, and the University of Dhaka, along with medical professionals from Pakistan and Malaysia.

EPIFF efficacy is assessed using classification accuracy (CA) and F1 score metrics. Furthermore, the receiver operating characteristic (ROC) analysis in terms of the True Positive Rate (TPR) and the False Positive Rate (FPR) is done to demonstrate the model's superiority in the classification task. CA measures the accuracy of the predicted samples over the total number of samples as modeled in Eq. 34

$$CA = \frac{TP+TN}{TP+FP+TN+FN} \tag{34}$$

Where $TP$ −True Positive; $TN$ −Ture Negative; $FP$ −False Positive; and $FN$ −False Negative.

Likewise, F1-Score accounted in terms of Precision ($Prec$) and Recall ($Rcl$) metrics. Herein, $Prec$ expressed in Eq. 35 quantifies the proportionality of correctly predicted samples over the number of predicted positive samples

$$Prec = \frac{TP}{FP+TP} \tag{35}$$

$Rcl$ (Sensitivity / TPR) measures the proportionality of the correctly predicted samples over the actual number of positive samples and is given in Eq. 36

$$Rcl = \frac{TP}{FN+TP} \tag{36}$$

Accordingly, the F1 score is presented in Eq. 37 and evaluates the model performance by successfully balancing the assessment across both positive and negative categories.

$$F1\ score = 2 \times \frac{Prec \times Rcl}{Prec \times Rcl} \tag{37}$$

The specificity investigates the model's ability to correctly reject the healthy samples and is presented in Eq. 38

$$\text{Specificity} = \frac{TN}{TN+FP} \tag{38}$$

The FPR is (1-specificity) investigates the model's reliability and accuracy by estimating the likelihood of incorrectly categorizing a negative sample as positive using Eq. 39

$$FPR = \frac{FP}{TN+FP} \tag{39}$$

To assess the degree of alignment between the model and the observed data, two additional metrics, namely the Mean Squared Error (MSE) and the $R^2$ scores are evaluated. The Mean squared error (MSE) is a common metric for evaluating the overall model accuracy by estimating the square difference between the actual and the model's predicted values as in Eq. 40

$$MSE = \frac{1}{n}\sum_{i=1}^{n} (Y_i - \hat{Y}_i)^2 \tag{40}$$

Where $n$ −number of samples; $Y_i$ −observed values; $\hat{Y}_i$ −predicted values.

Similarly, $R^2$ defined in Eq. 41 assesses the level of correlation between a model and the empirical data by estimating the difference between the squared sum residual with the total squared sum.

$$R^2 = 1 - \frac{\sum_{i=1}^{n} (Y_i - \hat{Y}_i)^2}{\sum_{i=1}^{n} (Y_i - \underline{Y})^2} \tag{41}$$

Where $Y_i$ −observed values; $\hat{Y}_i$ −predicted values; and $\underline{Y}$ −mean value.

### 3.2 Results and Discussion

To demonstrate the EPIFF-MARS accuracy in RD classification, an experiment is conducted on the NIH Chest X-ray 14 dataset. Accordingly, the individual confusion matrices for each class are formulated to facilitate a comparison between the model's predictions and actual outcomes, providing a detailed breakdown of its performance in terms of TP, TN, FP, and FN. The structure of the confusion matrix aligns true classes as row elements and predicted classes as columns, delivering a comprehensive overview of the model's effectiveness and is given in Fig. 4.

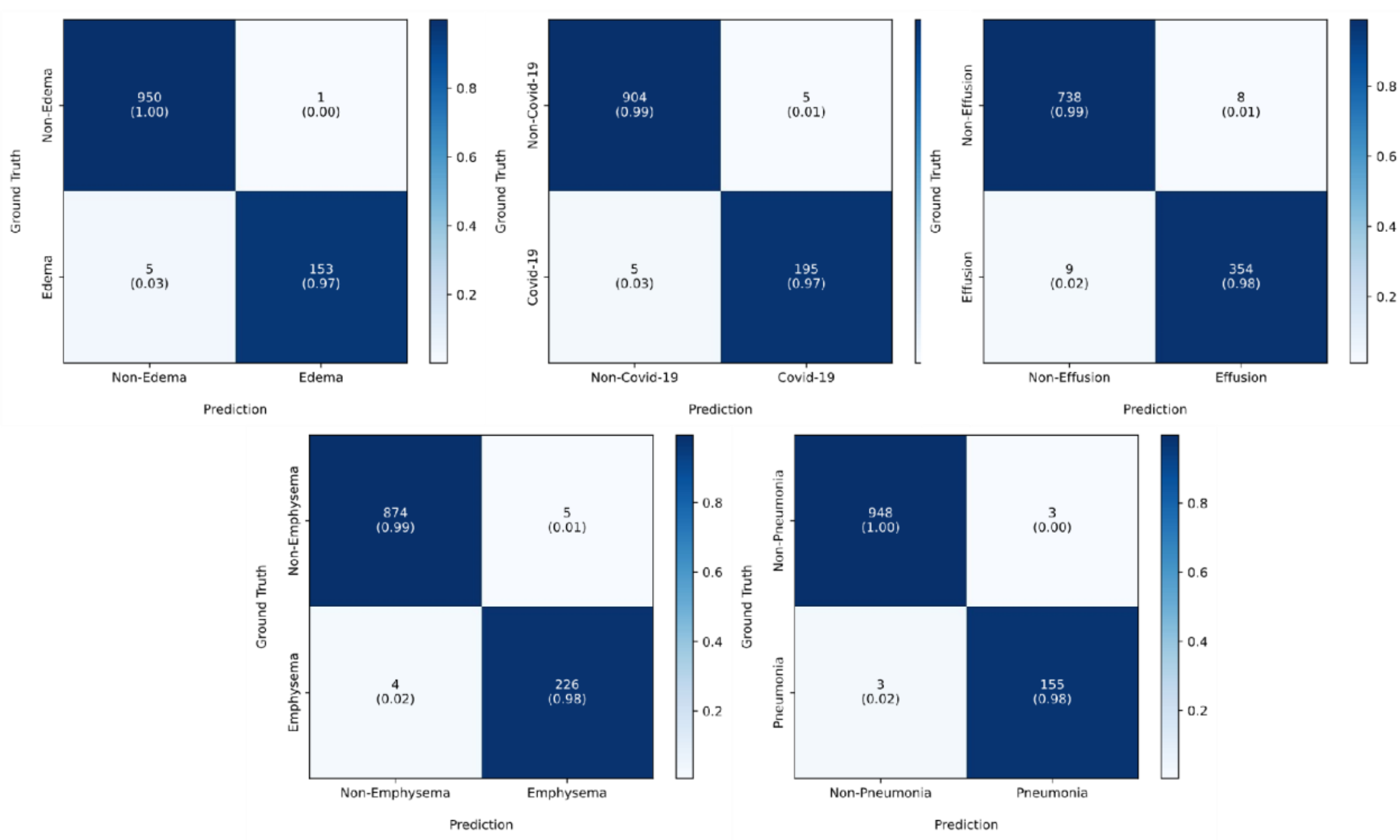

**Fig. 4 Confusion matrix of diverse RDs**

The confusion matrix in Fig. 4 indicates high TP and TN rates which are greater than 98% and 97% respectively showcasing its potential to differentiate between healthy and diseased patients across multiple categories, including Emphysema, COVID-19, Edema, Pneumonia, and Effusion. Further, achieving higher accuracy with imbalanced data is challenging due to the bias towards the majority class leads to increased FP and FN for the minority class. However, despite the imbalance, the EPIFF-MARS model achieved low rates of FP and FN which are lesser than 1% and 3% respectively demonstrating the extremely minimal misclassifications of EPIFF-MARS. The impact of TP, TN, FP, and FN is furthered by categorical evaluation of Accuracy, precision, sensitivity (Recall), specificity, and F1 Score of EPIFF-MARS in Table 4.

**Table 4. Qualitative Analysis of the EPIFF-MARS**

| Class Name | Accuracy | Precision | Recall | Specificity | F1 Score |
|---|---|---|---|---|---|
| **Edema** | 0.9946 | 0.9935 | 0.9684 | 0.9989 | 0.9808 |
| **Covid – 19** | 0.9910 | 0.9856 | 0.9789 | 0.9945 | 0.9822 |
| **Effusion** | 0.9847 | 0.9779 | 0.9752 | 0.9893 | 0.9765 |
| **Emphysema** | 0.9919 | 0.9784 | 0.9826 | 0.9943 | 0.9805 |
| **Pneumonia** | 0.9946 | 0.9810 | 0.9810 | 0.9968 | 0.9810 |

Upon examining Table 4, it is apparent that the EPIFF-MARS consistently exhibits superior outcomes, surpassing 97% in all metrics eventually demonstrating its effectiveness. Although, the 99% accuracy of EPIFF-MARS exhibits superiority perhaps misleads when dealing with imbalanced datasets, hence, the sensitivity(Recall) and specificity metrics are evaluated class-wise to register its reliability. The averaged percentage scores of 97.72 and 99.48 of sensitivity and specificity respectively in Table 4 are owed to the extremely low FP and FN rates as witnessed in Fig. 4. This achievement is due to the adaptation of a pairwise analysis of EPIFF-MARS for disease categorization which was particularly beneficial for dealing with imbalanced data that automatically qualifies the highly essential features representing each category despite the availability of significantly fewer samples. By focusing on these qualified features, EPIFF-MARS effectively handles the data imbalance. This illustrates the strong differentiating ability of EPIFF-MARS between healthy and diseased patients and contributes to improved patient care by minimizing misdiagnoses and ensuring timely treatment. Furthermore, achieving the averaged F1 score greater than 98% is particularly challenging and requires an intricate balance between the Precision and Recall offered by EPIFF-MARS, highlighting its potency in classification. These achievements are attributed to the highly distinctive fractal features that are effectively classified by the pairwise MARS classifier, which focuses on differentiating between specific pairs of classes by learning more nuanced decision boundaries. This potentially leads to improved accuracy compared to "one-vs-all", especially for complex medical datasets with similar classes. Also, EPIFF-MARS merits in terms of computational efficiency, improved accuracy, and interpretability warrant its extension to real-time implementation.

In addition, the relative analysis of EPIFF-MARS with existing state-of-the-art schemes (SOTA) is presented in Table 5.

**Table 5 Performance analysis of the proposed model with existing SOTA Schemes**

| Methods | AUC | Accuracy | Precision | Recall | Specificity | F1-Score |
|---|---|---|---|---|---|---|
| **EPIFF-MARS** | 0.9856 | **0.9912** | 0.9811 | **0.9764** | **0.9947** | 0.9787 |
| Red Deer (2023)[52] | - | 98.65 | 97.56 | 96.85 | - | 98.25 |
| XG-Boost-Beta–T (2023)[53] | - | 97.33 | 98.27 | 97.14 | 97.6 | 97.7 |
| ACPL (2022)[54] | 0.9436 | - | - | 0.7214 | - | 0.6223 |
| HealthyGAN (2022) [55] | 0.5600 | - | 0.5500 | 0.5500 | 0.5500 | 0.5500 |
| SA-CNN (2021)[56] | - | 96.67 | 96.69 | 96.67 | - | 96.67 |
| Kai et al (2022) [57] | **0.997** | 0.9796 | **0.9853** | 0.9738 | 0.9854 | **0.9795** |
| SRC-MT (2021)[58] | 0.9358 | - | - | 0.7147 | - | 0.6068 |
| DenseNet121(2017)[59] | 0.9928 | 0.929 | 0.8883 | 0.9793 | 0.8823 | 0.9307 |
| ResNet18 (2016)[60] | 0.9867 | 0.9509 | 0.9339 | 0.9686 | 0.9334 | 0.9504 |
| ResNet50 (2016)[60] | 0.9885 | 0.931 | 0.8951 | 0.9744 | 0.8903 | 0.9323 |
| Inceptionv3 (2016)[61] | 0.9907 | 0.9275 | 0.8866 | 0.9773 | 0.8805 | 0.9301 |
| VGG16 (2015)[62] | 0.9836 | 0.9274 | 0.9101 | 0.9439 | 0.9115 | 0.9264 |
| VGG19 (2015) [62] | 0.9878 | 0.9274 | 0.8883 | 0.9748 | 0.8834 | 0.929 |
| AlexNet (2012) [63] | 0.9791 | 0.9133 | 0.8827 | 0.9465 | 0.8826 | 0.9148 |

EPIFF achieved 98.56%, 99.12%, 98.11%, 97.64%, 99.47%, and 97.87% of AUC, accuracy, precision, recall, and F1 score respectively demonstrating its classification dominance over its peers. Specifically, it surpassed the recent DL schemes Red Deer[52], XG-Boost-Beta–T[53], ACPL[54], and HealthyGAN[55] thereby promising its effectiveness in acutely diagnosing abnormal samples. EPI's distinctive quality in effectively highlighting the low-contrast regions followed by localization using scale invariant fractals offers high class-wise structural discrimination with improved accuracy. However, EPIFF's performance lags slightly behind Kai et al [57], whose accuracy completely relied on the training of the complex Siamese deep neural network, demanding extensive hyperparameter tuning that is empirically derived to achieve optimality and hence, consumes intensive time. Rather, the elegant EPIFF requires less training and no parameter tuning, offering swift classification in fewer computations than its peers, making this model a promising tool for real-world applications in medical diagnostics.

In addition to the qualitative analysis, the Receiver Operating Characteristic (ROC) metrics namely Specificity (TPR) and Sensitivity (FPR) are plotted along the y and x-axis respectively for quantitatively analyzing EPIFF performance in Fig. 5. Generally, the model's ability to predict correct samples from the overall positive samples signifies its proximity to the top-left corner as shown in Fig. 5.

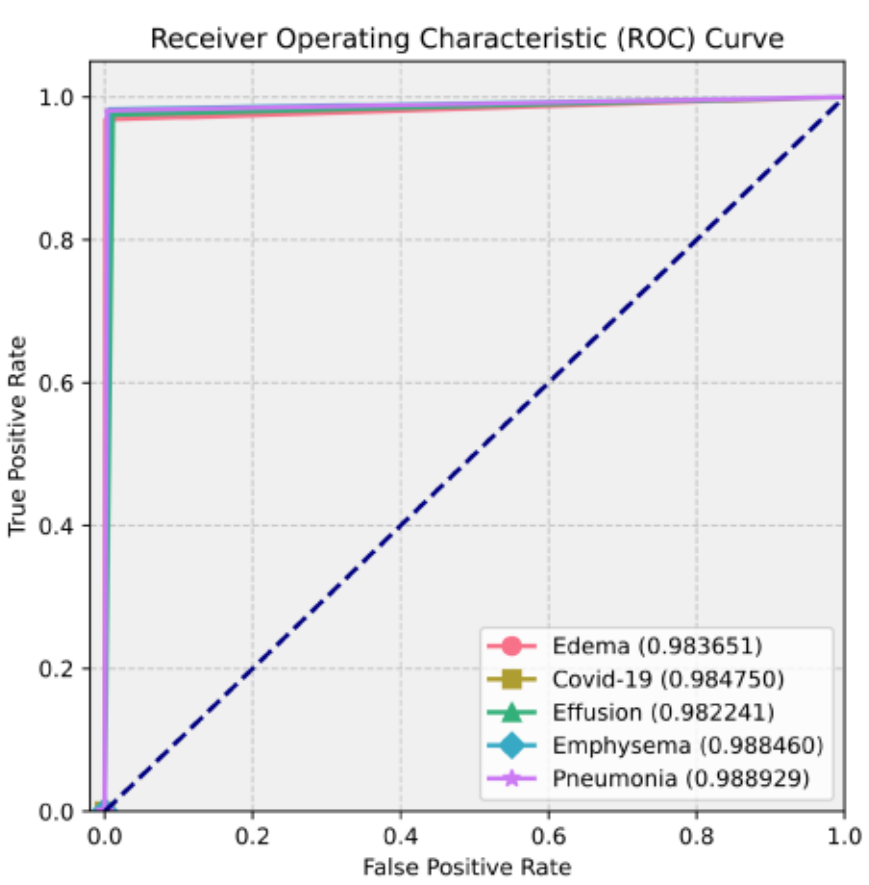

**Fig. 5 Proposed model's Receiver Operating Characteristics curve**

Fig. 5 illustrates the potential of EPIFF, emphasized by a substantial Area Under Curve (AUC). Notably, the AUC of ROC consistently exceeds 0.98 for all classes, indicating the superior classification ability of the EPIFF in distinguishing CXRs with and without abnormalities across all classes.

Consequently, Table 6 presents EPIFF-MARS performance with that of the recent competitors, assessed through the Area under the ROC Curves metric.

Table 6 Relative Analysis of Proposed Model's AUC in the ROC

| Methods | Edema | Effusion | Emphysema | Pneumonia |
|---|---|---|---|---|
| **EPIFF-MARS** | **0.9836** | **0.9822** | **0.9884** | **0.9889** |
| Hybrid CNN-T (2024)[64] | 0.8979 | 0.8839 | 0.9227 | 0.7651 |
| EEEA-Net-C2-KD(2024)[65] | 0.9006 | 0.8762 | 0.9147 | 0.7582 |
| SynthEnsemble (2023)[66] | 0.9103 | 0.8897 | 0.9294 | 0.7764 |
| Unichest (2023)[67] | 0.893 | 0.861 | 0.958 | 0.933 |
| FedKDF (2023) [68] | 0.8382 | 0.8298 | 0.8534 | 0.6609 |
| Nie et al (2023) [69] | 0.9000 | 0.9100 | 0.9400 | 0.8200 |
| BB-GCN (2023) [70] | 0.9100 | 0.9220 | 0.8970 | 0.7680 |
| SwinCheX (2022) [71] | 0.8510 | 0.8270 | 0.9140 | 0.7310 |
| LSAE (2022) [72] | 0.8401 | 0.8214 | 0.8547 | 0.7088 |
| ImageGCN (2022) [73] | 0.8800 | 0.8700 | 0.9200 | 0.7200 |
| Ouyang et al.(2021) [74] | 0.9000 | 0.8800 | 0.9400 | 0.7300 |
| Bose et al.(2021) [75] | 0.9210 | 0.8310 | 0.8610 | 0.7620 |
| Gundel et al. (2021) [76] | 0.8920 | 0.8850 | 0.9250 | 0.7650 |
| DGFN (2020) [77] | 0.8925 | 0.8751 | 0.9357 | 0.7791 |
| CheXGCN (2020) [78] | 0.8500 | 0.8320 | 0.9440 | 0.7390 |
| Baltruschat et al.(2019) [79] | 0.8460 | 0.8220 | 0.8950 | 0.7140 |
| DualCheXN (2019) [80] | 0.8520 | 0.8310 | 0.9420 | 0.7270 |
| ChestNet (2018) [81] | 0.8327 | 0.8114 | 0.8222 | 0.6959 |
| DNet D-161 (2018) [82] | 0.8880 | 0.8640 | 0.8980 | 0.7150 |
| AG-CNN (2018) [83] | 0.9240 | 0.9030 | 0.9320 | 0.7740 |
| CAN1 (2019) [84] | 0.8460 | 0.8280 | 0.8920 | 0.7210 |
| Wang et al. (2017) [49] | 0.8350 | 0.7840 | 0.8150 | 0.6330 |
| Yao et al. (2017) [85] | 0.8820 | 0.8590 | 0.8290 | 0.7130 |
| CheXNet (2017) [86] | 0.8870 | 0.8630 | 0.9370 | 0.7680 |
| DNet D-121 (2017) [86] | 0.8920 | 0.8850 | 0.9250 | 0.7650 |

The EPIFF-MARS model demonstrates superior performance in Table 6, achieving a higher Area Under the ROC Curve (AUC) compared to existing X-ray-based lung disease detection models in which most existing models rely on deep learning architectures, demanding significant computational resources while EPIFF-MARS is relatively simple. The superior achievements registered by the EPIFF-MARS are attributed to the effectiveness of the individual processing modules contributing significantly to the overall accuracy. Specifically, EPIFF outperforms recent methods [64-70] in discriminating various types of RDs, achieving significant improvements of 7% in edema detection, 8% in effusion detection, 5% in emphysema detection, and 22% in pneumonia detection. This efficacy is due to the pairwise analysis of the EPIFF-MARS approach that automatically identifies the most relevant features for each specific RDS category. Consequently, this leads to increased inter-class deviations, boosting classification accuracy while reducing the feature space.

Likewise, qualitative parameters estimated from CXRs for detecting COVID-19 are systematically compared with existing SOTA to illustrate the EPIFF efficacy in Table 7.

**Table 7 Relative Analysis of Proposed Model in COVID-19 Detection**

| Methods | Accuracy | Precision | Recall | Specificity | F1-Score |
|---|---|---|---|---|---|
| **EPIFF-MARS** | **0.9909** | 0.9785 | 0.9756 | **0.9945** | 0.9770 |
| **AIBF (2024)**[87] | 0.9583 | 0.9516 | 0.9621 | - | 0.976 |
| **ENet-B0-GAN (2023)**[88] | 0.841 | 0.86 | 0.86 | 0.946 | 0.856 |
| **Inception-v3 – GAN (2023)**[88] | 0.871 | 0.886 | 0.874 | 0.956 | 0.88 |
| ResNet50 (2023) [28] | 0.9567 | 0.9537 | 0.9594 | 0.9540 | 0.9566 |
| AlexNet (2023) [28] | 0.9362 | 0.9395 | 0.9334 | 0.9391 | 0.9364 |
| **ResNet-50 (2022)**[89] | 0.9614 | 0.963 | 0.96 | - | 0.963 |
| MKSC (2021)[90] | 0.9817 | **0.9813** | **0.9809** | 0.9825 | **0.9811** |
| CBAM (2021) [91] | 0.9633 | 0.9620 | 0.9530 | 0.9650 | 0.9650 |
| ECA-Net (2021) [92] | 0.9740 | 0.9761 | 0.9711 | 0.9770 | 0.9736 |
| **CheXnet (2020)** [93] | 0.9774 | 0.9661 | 0.9661 | 0.9661 | **0.9831** |
| **DenseNet201 (2020)** [93] | 0.9519 | 0.9506 | 0.959 | 0.9787 | 0.9504 |
| VGG-Net (2020) [94] | 0.9639 | 0.9632 | 0.9644 | 0.9640 | 0.9641 |
| DarkCovidNet (2020)[95] | 0.9632 | 0.9639 | 0.9621 | 0.9685 | 0.9630 |
| SE-Net (2018) [96] | 0.9693 | 0.9626 | 0.9734 | 0.9640 | 0.9680 |
| ResNet18 (2016) [60] | 0.9596 | 0.9610 | 0.9605 | 0.9624 | 0.9607 |

From Table 7, it is obvious that EPIFF-MARS outperformed other SOTA schemes in terms of the numerous ROC parameters. Specifically, its 99.10% accuracy signifies its dominance over its peers, whilst, it marginally falls behind the F1 scores of **CheXnet** [93]**. Despite this decline,** its superior consistency across the ROC spectrum is missing in this competitor. Likewise, MKSC [90] dominates EPIFF-MARS along the Precision and Recall dimensions thereby escalating the F1 score relatively. MKSC's dominance is mainly due to the exhaustive training nature of multi-kernel attention networks demanding extensive computations. Overall, the comparisons in Table 7 reveal the intense competition offered by EPIFF-MARS at a minimal complexity making it more suitable for real-time scenarios than the trending DL counterparts.

In addition, to prove the scalable nature of EPIFF-MARS, an extensive investigation with varying feature lengths of the NIH Chest X-ray 14 dataset is performed. To begin with, EPIFF-MARS extracts the essential features from CXRs and is subjected to MARS for disease classification. During training, by varying the threshold (maximum number of feature dimensions), MARS automatically selects the optimal feature subset within that range for each disease category. The achieved accuracy and F1 scores are presented in Table 8.

Table 8. Analysis of EPIFF-MARS model's Scalability

| Disease | Feature Size | | | |
|---|---|---|---|---|

| | Max Threshold Range | Selected Features | Accuracy | F1 Sscore | Execution Time (ms) |
|---|---|---|---|---|---|
| Edema | 50 | **30** | **0.9953** | **0.9883** | 30.71 |
| | 30 | 22 | 0.905 | 0.9055 | 22.51 |
| | 20 | 14 | 0.785 | 0.7943 | 20.96 |
| Covid-19 | 50 | **28** | **0.996** | **0.9899** | 37.65 |
| | 30 | 18 | 0.93 | 0.9346 | 32.93 |
| | 20 | 10 | 0.81 | 0.8288 | 29.42 |
| Effusion | 50 | **30** | **0.992** | **0.98** | 30.64 |
| | 30 | 26 | 0.94 | 0.9406 | 24.5 |
| | 20 | 14 | 0.754 | 0.7773 | 22.79 |
| Emphysema | 50 | **35** | **0.9933** | **0.9832** | 39.05 |
| | 30 | 26 | 0.915 | 0.9137 | 32.47 |
| | 20 | 15 | 0.76 | 0.800 | 20.76 |
| Pneumonia | 50 | **27** | **0.9893** | **0.9732** | 31.17 |
| | 30 | 20 | 0.965 | 0.9652 | 25.75 |
| | 20 | 11 | 0.825 | 0.8293 | 21.78 |

Table 8 showcases the EPIFF-MARS efficiency in feature selection wherein an averaged accuracy over 75% with a 30% feature reduction, and 99% with a 40% reduction is achieved at a significantly lower processing time under 20-30 ms thereby enabling faster and more resource-efficient classification tasks. This is due to the MARS quality in optimally selecting a small number of features through an intelligent selection process by focusing on highly relevant features and discarding unnecessary information. This leads to two key benefits: 1) Reduced complexity: The model is faster to train and requires less computational power, making it efficient for real-world applications. 2) Enhanced accuracy: By selecting the most informative features, MARS avoids irrelevant data, ultimately achieving superior classification results compared to models that rely on a larger generic feature set. Thereby, the intended model effectively maintains a trade-off between accuracy and efficiency.

Further, to justify the Robustness of the EPIFF-MARS, the model conducted the test using the balanced subset dataset composed from the ChestX-ray NIH 14 and the Covid Image Repository. This dataset included 1500 images, with 300 images for each disease category and the ROC accomplishments are tabulated in Table 9.

Table 9 Performance of EPIFF-MARS on the balanced dataset

| Disease | Confusion Matrix | | | | Accuracy | Precision | Recall | Specificity | F1 Score |
|---|---|---|---|---|---|---|---|---|---|
| | TP | FP | FN | TN | | | | | |
| Edema | 296 | 3 | 4 | 1197 | 0.9953 | 0.9900 | 0.9867 | 0.9975 | 0.9883 |
| Covid | 295 | 1 | 5 | 1199 | 0.9960 | 0.9966 | 0.9833 | 0.9992 | 0.9899 |
| Effusion | 294 | 6 | 6 | 1194 | 0.9920 | 0.9800 | 0.9800 | 0.995 | 0.9800 |
| Emphysema | 292 | 2 | 8 | 1198 | 0.9933 | 0.9832 | 0.9733 | 0.9983 | 0.9832 |
| Pneumonia | 294 | 4 | 6 | 1196 | 0.9933 | 0.9833 | 0.9800 | 0.9967 | 0.9833 |

Table 9 showcases the EPIFF-MARS consistent performance on a balanced dataset which is owed to the individual processing steps. Especially, the EPI transformation significantly enhanced the intensity of low-contrast regions in chest X-rays (CXRs), highlighting crucial areas that are more difficult to capture by the existing schemes demanding pre-processing. As a result, EPIFF-MARS registered average accuracies with specificities over 99% and averaged F1 scores with Sensitivity greater than 98% demonstrating its potential in the classification of diverse RDs. This achievement is particularly noteworthy compared to imbalanced datasets, suggesting the model's robustness.

Further, To comprehensively validate the effectiveness of the EPIFF-MARS pipeline, the process conducted an ablation study utilizing a balanced dataset that consists of a subset of 200 samples per category from the ChestX-ray14 dataset. Accordingly, the process adopted a two-step approach. Firstly, the model's performance is assessed directly on the raw data for performing baseline comparisons. Subsequently, the process systematically applied each transformation (EPI, Polar, and Fractal set) in EPIFF-MARS one by one to the raw data. Across each stage, the model's performance is evaluated to understand the impact of each processing stage in the EPIFF-MARS. The stage-wise outcomes of the developed EPIFF-MARS are presented in Table 10.

Table 10 Ablation study of EPIFF-MARS vs. Raw data

| **Image Features** | **Primary Class** | **Confusion Matrix** | | | | **Accuracy** | **F1 Score** | **Sensitivity** | **Specificity** |
|---|---|---|---|---|---|---|---|---|---|
| | | **TP** | **FP** | **FN** | **TN** | | | | |
| RAW | EDEMA | 128 | 115 | 72 | 685 | 0.813 | 0.5779 | 0.64 | 0.8563 |
| | COVID | 117 | 98 | 83 | 702 | 0.819 | 0.5639 | 0.585 | 0.8775 |
| | EFFUSION | 103 | 133 | 97 | 667 | 0.777 | 0.4725 | 0.515 | 0.8338 |
| | EMPHYSEMA | 95 | 177 | 105 | 623 | 0.718 | 0.4025 | 0.475 | 0.7788 |
| | PNEUMONIA | 100 | 125 | 100 | 675 | 0.775 | 0.4706 | 0.5 | 0.8438 |
| EPI | EDEMA | 141 | 94 | 59 | 706 | 0.847 | 0.6483 | 0.705 | 0.8825 |
| | COVID | 137 | 90 | 63 | 710 | 0.847 | 0.6417 | 0.685 | 0.8875 |
| | EFFUSION | 110 | 107 | 90 | 693 | 0.803 | 0.5276 | 0.55 | 0.8663 |
| | EMPHYSEMA | 144 | 112 | 56 | 688 | 0.8286 | 0.6316 | 0.72 | 0.8564 |
| | PNEUMONIA | 121 | 89 | 79 | 711 | 0.832 | 0.5902 | 0.605 | 0.8888 |
| EPI + Polar | EDEMA | 178 | 66 | 22 | 734 | 0.912 | 0.8018 | 0.89 | 0.9175 |
| | COVID | 182 | 45 | 18 | 755 | 0.937 | 0.8525 | 0.91 | 0.9438 |
| | EFFUSION | 142 | 41 | 58 | 759 | 0.901 | 0.7415 | 0.71 | 0.9488 |
| | EMPHYSEMA | 140 | 18 | 60 | 782 | 0.922 | 0.7821 | 0.7 | 0.9775 |
| | PNEUMONIA | 153 | 11 | 47 | 789 | 0.942 | 0.8407 | 0.765 | 0.9863 |
| **Proposed EPIFF (EPI + Polar + Fractals)** | EDEMA | 197 | 2 | 3 | 798 | **0.995** | **0.9875** | **0.985** | **0.9975** |
| | COVID | 196 | 1 | 4 | 799 | **0.995** | **0.9874** | **0.98** | **0.9988** |
| | EFFUSION | 194 | 5 | 6 | 795 | **0.989** | **0.9724** | **0.97** | **0.9938** |
| | EMPHYSEMA | 194 | 1 | 6 | 799 | **0.993** | **0.9823** | **0.97** | **0.9988** |
| | PNEUMONIA | 196 | 5 | 4 | 795 | **0.991** | **0.9776** | **0.98** | **0.9938** |

The results in Table 10 effectively highlight the impact of individual transformations involved in the EPIFF-MARS model against Raw data. As evident in Table 10, the processing across each stage contributes to significant improvement in the overall performance of the EPIFF-MARS model. Features extracted from the raw data registered an average accuracy of 78% with an F1-score of 49%, indicating limited effectiveness in differentiating abnormalities. In contrast, the extracted EPI transformed features achieved an average percentage of accuracy of 83.2 and an F1 score of 60.1 due to its contrast stretching nature that effectively highlights structural variations in low-contrast regions. Following EPI transformation, the features from the polar domain showcase significant improvement in ROC metrics due to its radial nature supplementing curved localization. Later, the extracted fractal features from the transformations raise the EPIFF-MARS performance significantly with an improvement of 7% in accuracy, 21% in F1 score, 23% in sensitivity, and 4% in specificity over EPI+polar features. This achievement is due to the exploitation of self-similar fractal patterns that effectively capture the structural variations present in diverse disease categories on chest X-rays. Thus, the ablation study showcases the potency of individual transformations formulating the EPIFF-MARS model.

Further, to investigate EPIFF's ability towards the noise, experiments are conducted involving the manual addition of various noise types namely Gaussian, Salt & Pepper, and Speckle to the CXRs, and their representations are illustrated in Fig. 6 with the averaged ROC outcomes stated in Table 11.

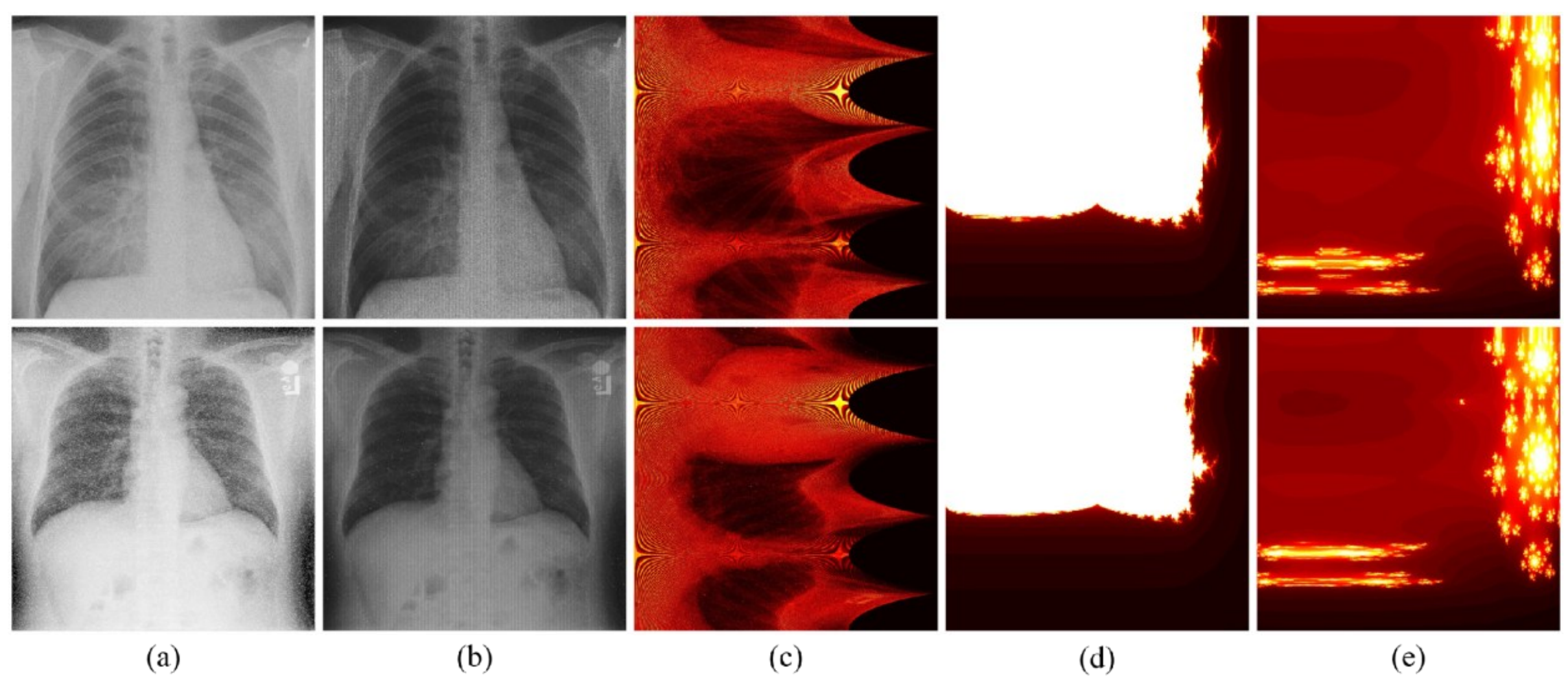

**Fig. 6 Sample noise added images and its representations: (a) noisy input image;(b) EPI transformed image; (c) Polar transformed image; (d) Mandelbrot Set; (e) Julia Set.**

**Table 11 Noise analysis of the EPIFF-MARS model**

| NOISE | Parameters | AUC | Accuracy | Precision | Recall | Specificity | F1-Score |
|---|---|---|---|---|---|---|---|
| **w/o Noise** | **-** | **0.9857** | **0.9906** | **0.9949** | **0.9939** | **0.9821** | **0.9944** |
| **Gaussian Noise** | **Variance (0.01, 0.1, 0.5)** | 0.9700 | 0.9789 | 0.9796 | 0.9600 | 0.9800 | 0.9697 |
| **Salt & Pepper Noise** | **Density (0.01, 0.1, 0.5)** | 0.9601 | 0.9500 | 0.9388 | 0.9583 | 0.9423 | 0.9485 |
| **Speckle Noise** | **Variance (0.01, 0.1, 0.5)** | 0.9549 | 0.9412 | 0.9184 | 0.9574 | 0.9273 | 0.9375 |

It is evident from Table 11, that the EPIFF produced consistent output regardless of diverse kinds of noise added with varying ratios. This is achieved due to the inherent smoothening nature of the introduced EPI which successfully averages out noise, thereby, resulting in diminishing its visibility in the image.

Further, to investigate the model's compatibility with the dataset, two additional metrics namely R2 score and MSE are estimated and presented in Table 12 indicating how well the model fits with datasets.

**Table 12 Analysis of the Prediction capability of the proposed model**

| Class Name | R2 Score | MSE |
|---|---|---|
| **Edema** | 0.9557 | 0.0054 |
| **Covid – 19** | 0.9390 | 0.0126 |
| **Effusion** | 0.9304 | 0.0153 |

| Emphysema | 0.9506 | 0.0081 |
|---|---|---|
| **Pneumonia** | 0.9557 | 0.0054 |

MSE values less than 0.01 in Table 12, emphasize the prediction ability of EPIFF-MARS as lower MSE signifies the closer alignment of predictions with actual results. In contrast, the R2 score surpassed 93% showcasing the model's generalization ability to fit with diverse datasets in terms of their prediction performance which is owed to the two-stage MARS process in formulating the classification model.

Furthermore, to illustrate the effectiveness of the EPIFF features, the experiments are performed using the extracted EPIFF features of CXR images coupled with diverse classification models, and the outcomes are tabulated in Table 13

**Table 13 Relative Analysis of the proposed model with other SOTA classification schemes**

| Model Setting | Accuracy | Precision | Recall | F1-Score |
|---|---|---|---|---|
| **EPIFF-MARS** | **0.9912** | **0.9811** | **0.9764** | **0.9787** |
| XGBoost | 0.8762 | 0.8810 | 0.8690 | 0.8750 |
| RandomForest | 0.9069 | 0.9480 | 0.8930 | 0.9197 |
| LightGBM | 0.9569 | 0.9480 | 0.9381 | 0.9430 |
| Deep Neural Network with Adam Optimizer | 0.9845 | 0.9783 | 0.9835 | 0.9809 |
| SVM | 0.9626 | 0.9410 | 0.9688 | 0.9547 |
| LogisticRegression | 0.8824 | 0.8840 | 0.8900 | 0.8870 |
| DecisionTree | 0.8320 | 0.8251 | 0.8144 | 0.8197 |

EPIFF coupled with MARS has surpassed its peers in Table 13 as witnessed in the registered performance scores. In comparison to a general linear regression or complex neural network, the MARS model is more flexible, computationally less expensive, and doesn't demand complex architecture. Moreover, MARS's well-maintained bias-variance trade-off with interpretability enables computing the predictors' functionality with their overall weightage. Overall it is claimed that EPIFF-MARS is superior in diagnostic accuracy with reduced complexity, is well suitable for less-experienced clinicians, and reduces interpretation variability among physicians. Also, the introduced methodology ensures that it is more adaptable to various RDs based on datasets and clinical contexts.

**4. Computational complexity**

To assess the practical efficiency and applicability of the EPIFF-MARS, a thorough analysis of its computational complexity, considering both time and space, has been conducted. The overall time complexity of the model is assessed by accounting for the time required at each developmental step. The model is partitioned into three stages. In the initial stage, the EPI transform is applied to an image ($I(n \times m)$) using a pixel-wise exponential function, incurring a time complexity of $O(n \times m)$ where $n \times m -$ represents the dimension of an image. Subsequently, in the second stage, the EPI is further transformed into fractal sets. The computational complexity for calculating Mandelbrot and Julia Sets is directly proportional to the number of iterations $(t)$ needed for each pixel, resulting in a complexity of $O(t(n \times m))$ from which features are extracted. Finally, the features are utilized in the MARS model for classification and its complexity is expressed as $O(B \times (N \times F) + M)$ where, $B -$ number of basis functions, $N -$number of samples in datasets, $F -$ total number of features applied to the model, and the additional '+M' term signifies the complexity associated with post-processing. Hence, the time complexity of the entire model is expressed in Eq. 41

$$T_{total} = O(n \times m) + O(t(n \times m)) + O(B \times (N \times F) + B)$$
$$T_{total} \approx O(B \times (N \times F) + B) \quad (42)$$

Further, to demonstrate the efficiency of the proposed model, the processing times (in seconds) are relatively compared with its peers and presented in Table 14.

**Table 14. Relative analysis of time complexity**

| Models | Model Processing Time (S) |
|---|---|
| ResNet-50 | 1349 |
| DenseNet-201 | 2399 |
| VGG-16 | 811 |
| DenseNet-169 | 2157 |
| Inception-v3 | 1239 |
| Simple CNN | 405 |
| VGG19 | 991 |
| ResNet-101 | 789 |
| **EPIFF-MARS** | **57** |

From Table 14 it is understood that there is a substantial variation in processing times across models, and notably, EPIFF-MARS stands out with an exceptionally short processing time, showcasing its high simplicity. In contrast, the other models demonstrate different training times, influenced by their respective architectures and complexities. The remarkable efficiency of the EPI-MARS model is attributed to less intricate developmental stages, making it a noteworthy option for tasks requiring swift classification.

Space complexity is assessed by considering each step in the realization of EPIFF-MARS. Upon processing the input image ($I(n \times m)$) by the EPI, yields a feature matrix of size whose complexity is $O(n \times m)$. Similarly, fractal sets generation demands $O(n \times m)$ space, as the resulting fractal set is in matrix form with dimensions matching the input image size. Finally, the space complexity of the MARS model is contingent on the number of basis functions ($B$), in addition to the space required for storing the features of samples in the dataset ends in $O(N \times F)$. Consequently, the total space requirement for the MARS model is to $O(B + (N \times F)$. Overall, the proposed model's space complexity is expressed in Eq. 42

$$S_{total} = O(n \times m) + O(B + (N \times F))$$
$$S_{total} \approx O(B + (N \times F)) \quad (43)$$

Thereby, the EPIFF-MARS is easy to implement on simple hardware structures and ensures low computational cost, making it suitable for use in resource-constrained settings.

**5. Conclusion**

In this paper, an elegant mathematical model for the automatic diagnosis of CXRs using a novel Exponential Pixelating Integral (EPI) is presented to classify various respiratory disorders for the benefit of both experienced medical professionals and novice physicians. The EPI transformation computes an exponential moving average over a set of pixels bound by an overlapping local 3x3 kernel. This technique effectively standardizes the images and helps to highlight all intensities, overcoming the limitations of differential intensity values in grayscale CXRs. The EPI is subsequently converted into Mandelbrot and Julia fractal representations via polar coordinates, enhancing the differentiation between normal tissue structures and abnormalities. The features from the three intermediaries are collated to construct an ensemble model of Multivariate Adaptive Regression Splines (MARS) for each pair of classes aiding Respiratory disorder classification. The rigorous analysis of the EPIFF-MARS on the large benchmark datasets demonstrated its consistency and superiority over its peers. Further, the noise analysis and Computational complexity revealed the robustness and amicability of EPIFF-MARS for diverse real-time scenarios. This study demonstrated a well-formulated design that can be used by the research community to develop AI solutions for diagnosing RDs. Although EPIFF-MARS effectively classified abnormality in CXRs, the acute identification of lesion positions needs further investigation for severity analysis thus motivating for deeper exploration of fractal-based textural features. Furthermore, optimizing EPIFF-MARS for clinical use requires extensive investigation by exploring model simplification and efficient algorithms to balance accuracy with computational efficiency. In addition, future efforts to broaden EPIFF-MARS to cover a wider range of RDs and explore its potential as a multimodal diagnostic tool, incorporating MRI, CT scans, and even ultrasound data alongside chest X-rays. Therefore, extending EPIFF-MARS to other imaging modalities will involve adapting it to the specific characteristics of each data type, potentially through data augmentation techniques or incorporating clinical settings into the feature extraction process.